\documentclass[reqno]{amsart}
\usepackage{oldlfont}
\usepackage{amssymb}
\usepackage{graphicx}

\voffset = - 1 cm
\hoffset = - 1.5 cm
\newtheorem*{ack}{Acknowledgment}

\newcommand{\bra}[1]{\langle #1 \! \mid}
\newcommand{\ket}[1]{\mid \! #1 \rangle}
\newcommand{\braket}[2]{\langle #1 \! \mid #2 \! \rangle}
\newcommand{\ev}[1]{\left\langle #1 \right\rangle}

\begin{document}
\title[The Shape of Spatial  Atoms]{Shape in an Atom of Space: Exploring quantum geometry phenomenology}
\author{Seth A. Major}
\date{May 2010}
\address{Department of Physics, Hamilton College,
Clinton NY 13323 USA}

\begin{abstract}
A phenomenology for the deep spatial geometry of loop quantum gravity is introduced.  In the context of a simple model, an atom of space, it is shown how purely combinatorial structures can affect observations.  The angle operator is used to develop a model of angular corrections to local, continuum flat-space 3-geometries.     
The physical effects involve neither breaking of local Lorentz invariance nor Planck scale suppression, but rather reply on only the combinatorics of $SU(2)$ recoupling.  Bhabha scattering is discussed as an example of how the effects might be observationally accessible.
\end{abstract}

\maketitle

\section{Introduction}

Quantum gravity phenomenology has developed into a broad field encompassing many possible effects arising from a more fundamental description of space-time. From cosmological perturbations to quantum decoherence to TeV scale black holes to particle astrophysics to violations of space-time symmetries, many, but not all, of the effects arise from the addition of a Planck scale. When the scale is at the naive Planck scale the effects are only observationally accessible with huge ``lever arms". For instance, when local Lorentz symmetry is violated the lever arm of bringing the Planck scale within reach of observation is the magnificent sensitivity of particle physics in the effective field theory framework to the breaking space-time symmetries \cite{jlm,limits,jlm_rev,lm_rev}. Cosmological distances can act as a level arm to raise effects from the additional scale into the realm of the observable even outside the effective field theory framework \cite{E_vary_c,fermi}. When the Planck scale is shifted due to the affect of additional physics from large extra dimensions \cite{larged} or a hidden gravitational matter sector \cite{hiddenm}, the level arm is more general, pulling the effective 4-dimensional Planck energy scale toward the natural scales of the standard model.

This paper introduces a model in which the lever arm is intrinsic to the discrete geometry of a spatial atom. As such it is an example of a phenomenology arising not from breaking local Lorentz invariance but rather from the structure of the fundamental description of space-time.  The model discussed here is based on the kinematic states of loop quantum gravity (LQG).

In this framework, an atom of spatial geometry is a single node of the graph which represents a quantum 3-geometry.  The combinatorics of the node determine geometric quantities, including angle \cite{angle} that is basis for the model discussed here.  From dimensional analysis the behavior of the angle spectrum does not depend on the Planck scale.  Rather, it is the combinatorics of the node that itself suggests a dimensionless ``shape" parameter.  This shape serves as an expansion parameter for the corrections to classical, flat-space continuum geometry. The model is based on the interplay between this combinatorics and the angle spectrum. The highly non-uniform spectrum gives rise both to an asymmetric distribution, which is parameterized by  the shape parameter, and to very large spins. While the shape involves no physical scales, the high spin determines an effective length via the quantum 3-volume. This length determines a mesoscopic scale above the Planck scale where the effects might be observable.  A simple analysis of Bhabha scattering is used as an example of how the effects might be accessible to observation.

In the remainder of this paper the angle operator is introduced in the context of the combinatorial framework of \cite{carlo_rev}. (The angle operator \cite{angle} in the embedded spin network context is reviewed in Appendix A.)  In section \ref{phenom} the details of the model are developed.  The example of combinatoric corrections in the context of Bhabha scattering is discussed  briefly in section \ref{scattering}. Comments on the model are collected in the final section. 

\section{Angle Operator}
\label{intro_angle}

The angle operator, originally defined in \cite{angle}, may be conveniently cast into the ``combinatorial framework" of loop quantum gravity \cite{carlo_rev} (see also \cite{bianchi,carlo_dual}).  The kinematics of this framework, relevant for spatial geometric operators, describes the space of spatial quantum 3-geometries. The state space,  the``combinatorial ${\cal H}$", is a separable Hilbert space, defined as equivalence classes of a direct sum of Hilbert spaces ${\cal H}_\Gamma$, each supported on a non-embedded, or abstract, (directed) graph $\Gamma$.\footnote{For  $L$ links and $N$ nodes the graph Hilbert space is 
${\cal H}_\Gamma = L^2[ SU(2)^L/SU(2)^{N-1}] $
where the Haar measure is used. This Hilbert space is used for the gravitation field operator. The equivalence relation is defined by automorphisms of the graph and by identifications induced by subgraph structure.  
See \cite{carlo_rev} for details.}

In the combinatorial framework the gravitational field operator, $L_l^i$ is the generator of the left $SU(2)$ action in ${\cal H}_\Gamma$ and has an interpretation as the flux of the inverse triad across the dual ``surface" of the link $l$.  Spin networks form a convenient basis on ${\cal H}_\Gamma$ in they are the eigenspace of spatial geometric observables. States in this spin network basis, $\ket{ \Gamma \, j_l \, v_n}$, are labeled by quantum ``numbers" consisting of the abstract graph $\Gamma$, the $SU(2)$ irreducible representations $j_l$ on the links $l$, and intertwiner labels $v_n$ for each node $n$ of the graph.  The intertwiner label $v_n$ has, in turn, an orthonormal basis labeled by a choice of a trivalent graph decomposition with a number of external links equal to the valence of the node $v$, and a set of $SU(2)$ irreducible representations on the internal links.

Non-embedded spin networks were first used by Penrose as a combinatorial foundation for 
Euclidean three-space \cite{penrose}.  Penrose \cite{penrose} and Moussouris  \cite{moussouris} constructed proofs that demonstrated that the angles of three-dimensional space could be modeled by operators on spin network states.  The kinematics of the combinatorial framework bring (this version of) loop quantum gravity into essentially the same framework used by Penrose and Moussouris.\footnote{The spin networks of \cite{moussouris} were all trivalent.  While this is also the case in the LQG context if one includes the sub-graphs of the intertwiners, to model spatial geometric quantities like angle and volume in LQG it is critical that the abstract graphs $\Gamma$ may contain higher valence nodes.}

In the combinatorial framework, the angle operator is defined on a node.  Links incident to $n$  are partitioned into three sets $C_{1}$, $C_{2}$, and $C_3$.  (One may visualize the partitioning as arising from three regions in the surface dual to the node, as represented in figure \ref{surfaces3}.  However, the combinatorics only requires a tripartite partition.) Three gravitational field operators $L^i_{1}$, $L^i_{2}$, and $L^i_{3}$ are associated to these partitions. For instance, if there are $s_1$ links $l_{1_m}$, $m=1,\dots,s_1$, in the partition $C_1$ then $L^i_1 = \sum_{m=1}^{s_1} L^i_{1_m}$.    In terms of these field operators the quantum angle operator between dual surfaces corresponding to partitions $C_1$ and $C_2$ is
\begin{equation}
	\label{Aop}
\hat{\theta}_{(12)} := \arccos \frac{L^i_{1} L^i_{2}}
{|L_{1}| \, | L_{2} |},
\end{equation}
in which $|L| = \sqrt{L^{2}}$.  Because the partitions are exhaustive and because of gauge invariance, $\sum_{k=1}^3 L^i_k =0$. This partitioning of links incident to $n$ gives a preferred (class of) intertwiners $v_n$.  These are given by three, trivalent tree graphs that connect in an ``intertwiner core".  I will label the links of the intertwiner core with irreducible representations $j_k$ (and, later with $n_k = 2j_k$) and the basis of the intertwiner core by $\ket{j_1 \, j_2 \, j_3}$ (later by $\ket{\vec{n}}$).  For the purposes of the eigenvalues of the angle operator, the remaining labels on the internal edges are not important.  However, they do play a role in the phenomenology discussed in the next section.

Deriving the spectrum of the angle operator of equation (\ref{Aop}) is a simple exercise in angular momentum algebra  \cite{angle}
\begin{equation}
\label{angle_spectrum}
\begin{split}
\hat{\theta}_{(12)} \ket{j_1 \, j_2 \, j_3} &= \theta_{(12)} \ket{j_1 \, j_2 \, j_3}  \text{ with } \\
\theta_{(12)} &= {\rm arccos} \left( \frac{j_{3}(j_{3}+1) -
j_{1}(j_{1} +1) - j_{2}(j_{2}+1)} {2 \left[ j_{1}(j_{1} +1 ) \, 
j_{2}(j_{2} +1 ) \right]^{1/2}} \right).
\end{split}
\end{equation}
 
The original idea of Penrose was to measure angle via correlations between two disjoint sets of links \cite{penrose}.  As noted briefly above in the combinatorial framework it is convenient to visualize the action of the angle operator on  the dual surfaces of links in the three partitions. Thus, the closed dual surface of the node, topologically $S^{2}$, is partitioned into three regions, $S_{1}$, $S_{2}$, and $S_{3}$, such that all the regions $S_{k}$ are simply connected and such that the annular region $S_2$ separates $S_1$ and $S_3$, as shown in figure \ref{surfaces3}. From this picture the angle defined above is represents the zenith or angle $\theta$ of spherical coordinate systems. As the partitions or selection of regions $S_k$ is varied, the possible core intertwiner labels vary, changing the spectrum of the angle. Although this picture is convenient and is also how the embedded version of the operator is defined, this construction is not necessary for the combinatorics.  All that is required in the combinatorial setting is the partition of the incident links.  One may also use a more symmetric definition of the three surfaces, as shown in figure \ref{surfaces3}(a.).

\begin{figure}
        \includegraphics[width=4in]{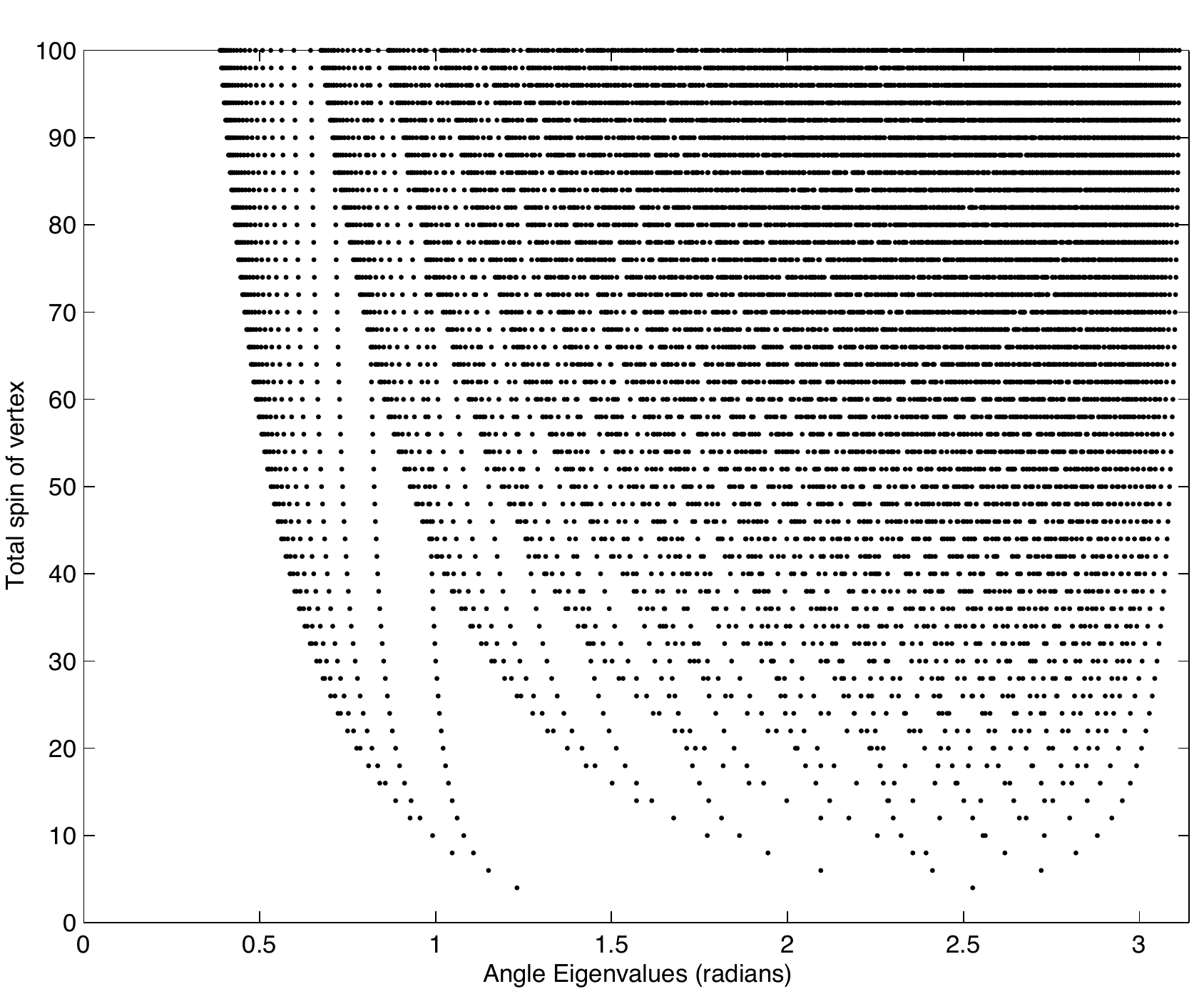}
    \caption{\label{ang_spec} Angle operator spectrum for increasing flux at a node, or ``total spin" of the vertex  $n=2\sum_k j_k$.  The complete spectrum is plotted for total spins from 3 to 100. With few eigenvalues at small angle and the non-uniform spacing, the spectral distribution differs strongly from the continuum distribution. From \cite{mikes,major_seifert}.}
\end{figure}

As is clear from a glance at the spectrum, figure \ref{ang_spec}, there are two aspects of the distribution of angles in the continuum that are hard to model.  First, small angles are sparse.  Second, the distribution of values is asymmetric and weighted toward large angles.  As discussed in \cite{mikes,major_seifert} the asymmetry persists even when the spins are large.  The effects discussed in this paper are due to this asymmetric bias.\footnote{The striking fan-like structure in the spectrum is discussed in \cite{mikes}.}

In sum, the angle operator is simply defined in the combinatorial framework of \cite{carlo_rev}.  The angle operator of equation (\ref{Aop}) acts on nodes and the spectrum may be expressed in terms of the $SU(2)$ representations  of the intertwiner core at a single node,  determined by a partition of the links incident to the node.  This angle operator is closely related to the combinatorial operator discussed in the works of Penrose \cite{penrose}. Finally, as was done originally in \cite{angle} the angle operator may also be defined in terms of the electric flux variables and the usual embedded graphs of LQG.  This is discussed in Appendix A, along with some further comments on the definition of the angle operator in that framework.

The notation for the remainder of the paper is as follows. Twice the sum of the representations on the links incident to the node in partition $C_k$ is denoted by the ``flux"  $s_k$ also denoted $\vec{s}$.  In the dual surface picture this is the flux of spin through the respective surfaces.  In the literature this is also sometimes called ``area" (see, for instance, \cite{u1}).   The quantities $n_k= 2 j_k$ uniquely specify the intertwiner core that  
``collect'' the fluxes $s_k$ from each of the three partitions $C_k$, or dual surfaces.  The relevant orthonormal states are $\ket{n_1 \, n_2 \, n_3}$, denoted $\ket{\vec{n}}$.The fluxes $s_k$ and core labels $n_k$ are distinct and satisfy $n_k \leq s_k$. In terms of labels $n_k$  the angle becomes
\begin{equation} \label{angformn}
\theta_{12} = \arccos \left( \frac{n_3 (n_3 + 2) - n_1 (n_1 + 2) - n_2 (n_2 + 2)}
{2 \sqrt{n_1(n_1 + 2)n_2(n_2+2)}} \right).
\end{equation}

\section{Combinatorial phenomenology}
\label{phenom}

The semi-classical, or continuum limit of the angle operator was numerically investigated in \cite{mikes,major_seifert}. To model an atom of 3-geometry we made the ansatz that the probability measure on the space of intertwiners was uniform, that every possible set of labels on internal links of the node was equally likely. We further made a simplifying assumption that all incident links to the node were spin-$\tfrac{1}{2}$, simple and monochromatic.  This assumption made the combinatorial problem quite tractable. 

In the simple monochromatic case the dimension of the Hilbert space of the node is the number of intertwiners at a fixed flux $\vec{s}$ and intertwiner core $\vec{n}$ , $\text{dim} {\cal H}_n = \text{dim} {\mathcal H}_{j_1,..,j_{s_1},j_0}$.
This equals the product of the distinct ways of labeling the three branches of the intertwiner graph. For each branch $k$ the dimension of the intertwiner space for fixed flux $s_k$ and core label $n_k$ is equivalent to a well-known path counting problem \cite{mikes}. The result for one branch $k$ is \cite{mikes}
\begin{equation} \label{collnum}
Q(s_k, n_{k}) = \frac{n_k+1}{s_{k}+1} {s_{k}+1 \choose 
\frac{n_{k}+s_k}{2} + 1}.
\end{equation}
With the assumptions of uniform probability and simple monochromatic nodes, the combinatorics gives a probability distribution. Since each branch contributes a factor as in equation (\ref{collnum}), $Q(\vec{s},\vec{n}):= \prod_{k=1}^3 Q(s_k,n_k)$.  The probability distribution is then 
$
p_{\vec{s}}(\vec{n}) = Q(\vec{s},\vec{n})/|Q(\vec{s},\vec{n})|
$
where the norm $|Q(\vec{s},\vec{n})| = \sum_{\vec{n}} Q(\vec{s},\vec{n})$, i.e. is the dimension of the invariant intertwiner with fluxes $\vec{s}$.

It was apparent in the numerical studies of \cite{mikes} that the non-uniformity in the spectrum shifted the probability distribution $p_{\vec{s}}(\vec{n})$ away from the usual $\sin \theta$ distribution of angles in three dimensional flat space. To recover this it was necessary to take large fluxes, corresponding to a very high valence node, and, in particular $1 \ll s_j \ll s_3$, $j=1,2$; the ``background geometry",  $s_3$, must be robust. I'll call fluxes $\vec{s}$ that satisfy these relations ``semi-classical fluxes".

There is another reason why the we might wish to consider nodes with large spin.  Most physical processes we currently consider, such as scattering events, are ``local" on the scale of the theory being tested. But in terms of the quantum geometry the scales are very large, typically many orders of magnitude above the Planck scale.  In the volume operator likely to be relevant for the combinatorial framework, the volume scales as the $(\text{total flux})^{3/2}$.\footnote{See section \ref{concl} for a discussion on volume on LQG.} The scaling with volume can be used to define an effective length $\ell_s$ and an effective energy $M_s = M_{Pl}/\sqrt{s}$. A surprising result in \cite{mikes,major_seifert} suggests that to model the correct distribution of angles in 3-space, the total fluxes were $10^{32}$ giving an effective length scale of  about $10^{-19}$ m, a perhaps not altogether hopeless scale.

These initial results suggest that states with semi-classical fluxes are a promising source for phenomenology. The remainder of the paper focuses on this model of the atom of quantum 3-geometries: simple monochromatic nodes with uniform probability distribution on the intertwiners and semi-classical fluxes.  (While it is expected that the simple monochromatic node will dominate the sum, relaxing this assumption will change the quantitative results reported here.\footnote{It seems likely that the generalization of the methods of \cite{u1} can give the general case.}) The following analysis shows that the degree to which the semi-classical flux relations are satisfied determines the size of the combinatorial corrections investigated here.
 
The combinatorics of the model can be solved analytically for semi-classical fluxes. For large flux $s$ the normalized probability distribution of $Q(s, n)$ 
is given by \cite{mikes}
\begin{equation}
\label{dist}
\frac{n+1}{s+1} \exp\left[ - \frac{n^2 +2n}{2(s+1)} \right] \simeq \frac{n}{s} \exp \left( -\frac{n^2}{2s} \right) =:  P_{s}(n)
\end{equation}
Interestingly, the distribution $P_{s}(n)$ is the Rayleigh distribution for the distance $n$ covered in $2s$ steps in an isotropic random walk with unit step size in {\em two} spatial dimensions.  Since each branch of the intertwiner is independent, the distribution for the whole intertwiner is simply the product
\begin{equation}
\label{ndist}
p_{\vec{s}}(\vec{n}) \simeq \prod_{i=1}^3 \frac{n_i}{s_i}  \exp \left( -\frac{n_i^2}{2s_i} \right)
\end{equation}
The distribution $P_s(n)$ is peaked at $\sim \sqrt{s}$ and has a width (`FWHM') of approximately $2 \sqrt{ s \ln( 2) }$.  For large $s_k$ the likely values of $n_k$ are also large and we can approximate the $\vec{n}$ by continuous values.  In this case $\cos \theta$ becomes, from equation (\ref{angformn}),
\begin{equation}
\label{largenform}
\cos \theta \simeq \frac{n_3^2 - n_1^2 - n_2^2}{2n_1n_2} \text{  or,  } \theta(\vec{n}) = \arccos \left(  \frac{n_3^2 - n_1^2 - n_2^2}{2n_1n_2} \right).
\end{equation}

To study the effects of the combinatorics it is useful to work with the exact, discrete quantum states before the continuum approximation.
The states of the spatial atom are labeled by the full intertwiner $v_{n}$.  However, accessible measurements of the atom include 3-volume, (roughly) determined by the total flux, and angle, determined by the states $\ket{\vec{n}}$ of the intertwiner core.  In this model the fluxes $\vec{s}$ determine a mixed state,
\begin{equation}
\label{mixedstate}
\rho_{\vec{s}} = \sum_{\vec{n}} p_{\vec{s}}(\vec{n}) P_{\ket{\vec{n}}}
\end{equation}
where $P_{\ket{\vec{n}}}$ is the projector on the orthonormal basis of the intertwiner core. The sum is over the admissible 3-tuple of integers $\vec{n}$ such that $n_i \leq s_i$. In the discrete case the projector is $P_{\ket{\vec{n}}} = \ket{\theta_I} \bra{\theta_I}$, as usual, where the orthonormal $\ket{\theta_I} = \sum_{\vec{n}} c_{\theta_I} (\vec{n}) \ket{\vec{n}}$. At a fixed angle the amplitudes $c_{\theta_I} (\vec{n})$ vanish except when $\vec{n}$ gives $\theta_I$. Due to the symmetry of the angle operator, angles enjoy a degeneracy under the exchange of $n_1$ and $n_2$.  It would be interesting to explore possible effects of the relative phases in $c_{\theta_I} (\vec{n})$, but they will play no role in the following.

The probability of finding the angle eigenvalue $\theta_I$ in the mixed state $\rho_{\vec{s}}$ is 
\begin{equation}
\label{discrete_dist}
\text{Prob}(\theta = \theta_I ; \rho_{\vec{s}}) = \text{tr}\left( \rho_{\vec{s}} P_{\theta_I} \right)
= \sum_{\vec{n}} p_{\vec{s}}(\vec{n}) |\braket{n}{\theta_I} |^2 \equiv
p_s(\theta).
\end{equation}
This procedure can be used to calculate $p_{\vec{s}} (\theta)$ in the continuum approximation.

In the continuum the mixed states for nodes with fixed semi-classical fluxes have density matrix
\begin{equation}
\hat{\rho}_s = \int d^3n P_{\vec{s}}(\vec{n}) \hat{P}_{\ket{\vec{n}}}
\end{equation}
where $\hat{P}_{\ket{\vec{n}}}$ is the projector on the states $\ket{\vec{n}}$.  So for large fluxes and a value of the measured angle $\theta$, now taking continuous values, within an interval $\Delta \theta = (\theta-\delta \theta,\theta +\delta \theta)$ the geometric probability distribution is 
\begin{equation}
\text{Prob}(\theta \in \Delta \theta; \hat{\rho}_s) = \text{tr} \left( \hat{\rho}_s \hat{E}_{\Delta \theta} \right) = \int d^3n P_{\vec{s}}(\vec{n}) \bra{\vec{n}} \hat{E}_{\Delta \theta} \ket{\vec{n}}
\end{equation} 
where $\hat{E}_{\Delta \theta}$ is the projector onto the interval $\Delta \theta$. Geometrically it projects the state onto a (thickened) surface in $\vec{n}$-space given by $\theta(n) \in \Delta \theta$.  Taking the limit $\delta \theta \rightarrow 0$ gives, heuristically, the geometric probability distribution 
\begin{equation}
\label{rawdisttheta}
P_{\vec{s}}(\theta)  := \int d^3n  \, p_{\vec{s}}(\vec{n}) |c_\theta(\vec{n})|^2 \delta \left( \theta - \theta(n) \right). 
\end{equation}
This is the continuum approximation to equation (\ref{discrete_dist}). The normalization of the continuum approximation $\ket{\theta}$ states is determined by the area of the surface $\theta = \theta(\vec{n})$.  This gives $|c_\theta(\vec{n})|^2 =  |c_\theta(\vec{s})|^2$ and so $|c_\theta(\vec{n})|^2$ becomes an overall factor in the above integration.

The integration of equation (\ref{rawdisttheta}) is straightforward and done in Appendix B.  The key step in the calculation is the identification of the ``shape parameter" $\epsilon := \sqrt{s_1 s_2}/{s_3}$, which measures the asymmetry in the distribution of angles.  Small for semi-classical fluxes, $\epsilon$ is the parameter used for the expansion of the combinatorial corrections. 

\begin{figure}
      \begin{center}
	\includegraphics[scale=1.0]{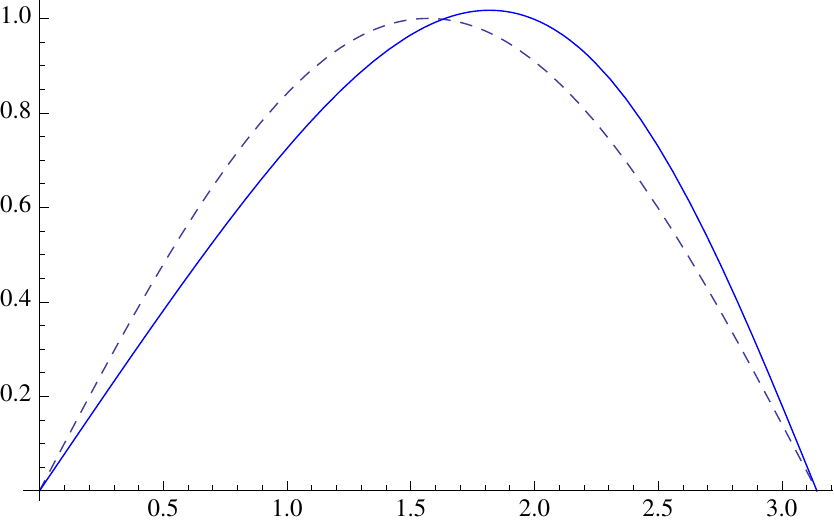}
      \end{center}
\caption{\label{dist_plot} The distribution of angles from combinatorial geometry $\rho_\epsilon (\theta)$ (red) compared to the usual $\sin \theta$ measure (blue). The distribution contains corrections to 4th order in $\epsilon$  with $\epsilon = 0.1$.}
\end{figure}

\begin{figure}
      \begin{center}
	\includegraphics[scale=1.0]{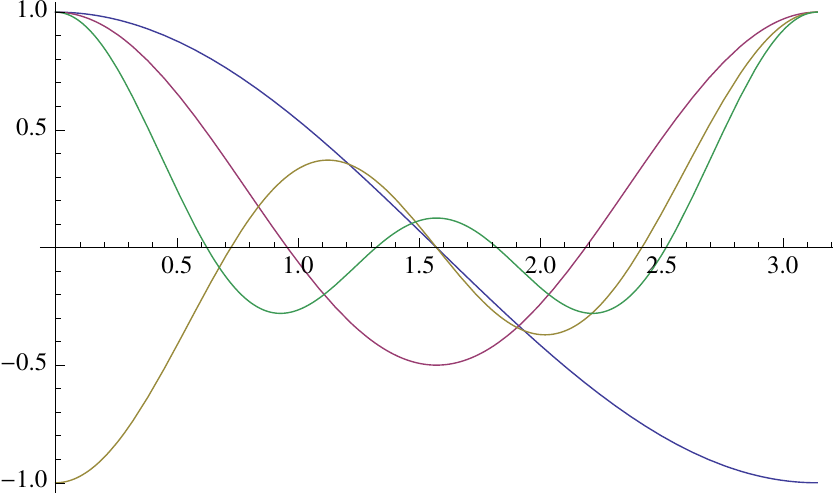}
      \end{center}
\caption{\label{Q_corrections}  The functions of $\rho_\epsilon(\theta)$ to 4th order in $\epsilon$ normalized to 1 (blue,red, green, yellow).}
\end{figure}

The result gives a modified distribution of polar angles $\theta$ given by $\rho_\epsilon(\theta) = P_{\vec{s}}(\theta)/N$. As shown in equation (\ref{normN}) the normalization $N$ is determined by requirement of recovering the continuum distribution in the limit of vanishing $\epsilon$.  The resulting distribution or measure, when expressed in terms of Legendre polynomials and to $O(\epsilon^3)$, is 
\begin{equation}
\rho_\epsilon (\theta) \simeq \sin \theta \left( 1 - \frac{8}{\pi} P_1(\cos \theta) \epsilon  + \frac{3}{2} P_2(\cos \theta) \epsilon^2 \right)
\end{equation} 
from equation (\ref{cgdist}). The affect of the change is that the `shape' of space is altered by the combinatorics of the vertex; the local angular geometry differs from flat Euclidean 3-space. For instance, the expectation value of an angular quantity $f(\theta)$ in the mixed state $\rho_{\vec{s}}$ is corrected
\begin{equation}
\ev{f(\hat{\theta})}_s = \int d \theta \text{ tr} \left( \hat{\rho}_s \hat{E}_{\Delta \theta} \right) = \int d \theta  f(\theta) \rho_{\epsilon} (\theta).
\end{equation} 
The distribution reproduces the usual distribution of angles in the limit of vanishing shape parameter.  This is an analytic expression of what was found numerically in \cite{mikes} and is a manifestation of the spin geometry theorem.  The $\rho_\epsilon(\theta)$ distribution is compared to the usual one in figure \ref{dist_plot}.  The angular corrections are shown in figure \ref{Q_corrections}. 

In the semi-classical flux limit, the values of the fluxes $\vec{s}$ enter into the distribution only through the shape parameter.  One can average over semi-classical fluxes, and thus $\epsilon$, which effectively determines an average shape parameter $\epsilon$.  

As a result of the angle spectrum and the uniform probability measure on the intertwiner space, combinatorial effects of the toy model are parameterized by a single dimensionless shape parameter $\epsilon = \sqrt{s_1 s_2}/s_3$.  While these effects would be in principle observable at any flux, the results here are valid for semi-classical flux,  $1 \ll s_j \ll s_3$ for $j=1,2$.  In this model the total flux $s = \sum_i s_i$ determines the 3-volume of the spatial atom and thus an effective length scale, $\ell_s = \sqrt{s} \ell_P$, greater than the fundamental discreteness scale of $\ell_P$. So while the the shape parameter $\epsilon$ is free of the Planck scale, the effective length scale, determined by total flux $s$, is tied to the discreteness scale of the theory.

\section{Example: Scattering}
\label{scattering}

If the scale $\ell_s$ of the spatial atom is large enough then the underlying geometry would be accessible to observations of particle scattering. To see how the combinatorial effects might be manifest I'll briefly discuss  combinatoric corrections to Bhabha scattering. This process is convenient because the $e^+e^-$ scattering process involves ``point-like" fundamental particles and for the practical reason that the data is readily available \cite{bhabha}. This serves as an example of possible combinatorial corrections and will not yield constraints on the model parameters.  In the experiment reported on in \cite{bhabha} the center-of-mass energy was 29 GeV, corresponding to a rough length scale of $10^{-17}$ m in the center-of-mass frame.  This length scale corresponds to a flux of roughly $10^{36}$ so combinatorial corrections should be negligible.  Nevertheless the data serves as a simple example of how $\epsilon$ might be constrained using a more complete analysis. 

The pure QED differential cross section for the process at lowest order is \cite{bhabha}
\begin{equation}
s \left( \frac{d \sigma}{d \Omega} \right)_{\text{QED}} =  \frac{\alpha^2}{4} \frac{(3+\cos^2\theta)^2}{(1-\cos \theta)^2}
\end{equation}
where $s$ here is the square of the center of mass energy.  The differential cross section at 29 GeV is affected by electroweak effects.  To lowest order the effect is roughly to reduce the differential cross section by 1-2 \% \cite{bhabha}.  As discussed below the combinatorial corrections mimic this correction. 

Since this is a purely kinematic model, I assume that the observation of the particle shower and subsequent reconstruction of the scattered angle is simply a measurement of angle.  There are (at least) two effects of the discrete geometry. The local geometry and angular distribution on small scales and averaging over angles are modified.  The former is dominant.

The probability distribution of angles effectively alters the local angular geometry and leads to an angle-dependent rescaling of the cross section.  The scattering data is binned in terms of the solid angle.  The number of events $N_i$ counted in an interval of $\theta$ is proportional to the differential scattering cross section 
\begin{equation}
s \left( \frac{d \sigma}{d \Omega} \right)_{i} \propto \frac{N_i}{\Delta \Omega_i} 
\end{equation} 
To account for the asymmetry in the combinatorics, or equivalently the modification of the local angular geometry of space, the angular normalization must be adjusted.  The solid angle is modified $\Delta \omega \rightarrow \rho_\epsilon (\theta) d\theta d\phi \equiv Q_\epsilon (\theta) d \Omega$. This is the first and dominant effect.

The second effect arises in the averaging an angular quantity $f(\theta)$ such as the differential cross section.  The angle is only measured to some finite precision so the quantities are averaged over an interval $\Delta \theta = (\theta_o - \delta \theta,\theta_o + \delta \theta)$,
\begin{equation}
\overline {f(\theta)} = \frac{ \int_{\Delta \theta} \rho_\epsilon (\theta) f(\theta) d \theta }{ \int_{\Delta \theta} \rho_\epsilon (\theta) d \theta }
\end{equation}
Expanding the function gives a weighted Taylor series for the average
\begin{equation}
\label{coraveexp}
\overline {f(\theta)} \simeq f(\theta) + f'(\theta) w_1(\theta,\delta \theta, \epsilon) + \tfrac{1}{2} f''(\theta) w_2(\theta,\delta \theta, \epsilon)
\end{equation}
where the weights are given in Appendix B.  As the leading corrections are $O(\delta \theta^2)$ or $O(\delta \theta^2 \epsilon)$, the effects are negligible for an experiment in which $\delta \theta$ is about 0.002 rad.
Thus the comparison will be only from effects arising from the asymmetry in the local geometry arising from the combinatorics of the angle operator. 

Short distance modifications to QED may be useful expressed in the Drell parameterization \cite{drell,bhabha}
\begin{equation}
\left( \frac{d \sigma}{d \Omega} \right) / \left( \frac{d \sigma}{d \Omega} \right)_{QED}  = 1 \mp \left(\frac{3 s}{\Lambda^2_\pm} \right) \frac{ \sin^2 \theta }{3 + \cos^2 \theta} 
\end{equation}
These correspond to a short range potential added to the Coulomb potential. In this model the combinatorics of space is fixed; the excitations of the geometry only occur at much higher energy scales.  The local, discrete geometry manifests itself through the combinatorial corrections to the local geometry.  The shape corrections would be evident above the energy scale $M_s$ and give an angle-dependent scaling of the cross section
\begin{equation}
\frac{d \sigma}{d \Omega} \rightarrow \frac{d \sigma}{d \Omega} Q_\epsilon^{-1}(\theta) = \frac{d \sigma}{d \Omega} \left( 1 + \frac{8}{\pi} \cos (\theta) \epsilon + \dots \right).
\end{equation}
The short distance, shape corrected cross section has a correction of the form
\begin{equation}
\left( \frac{d \sigma}{d \Omega} \right) / \left( \frac{d \sigma}{d \Omega} \right)_{QED} = 1 \mp \left(\frac{3 s}{\Lambda^2_\pm} \right) \left( \frac{ \sin^2 \theta }{3 + \cos^2 \theta} \right) \left( 1 + \frac{8}{\pi} \cos (\theta) \epsilon + \dots \right)
\end{equation}
It remains to be seen whether a derivation along the lines of \cite{drell} that incorporates shape corrections would yield this form of the correction.  Nevertheless, this example serves to show that scattering processes could constrain the model parameters. 

\begin{figure}
      \begin{center}
      \begin{tabular}{c}
	\includegraphics[scale=0.65]{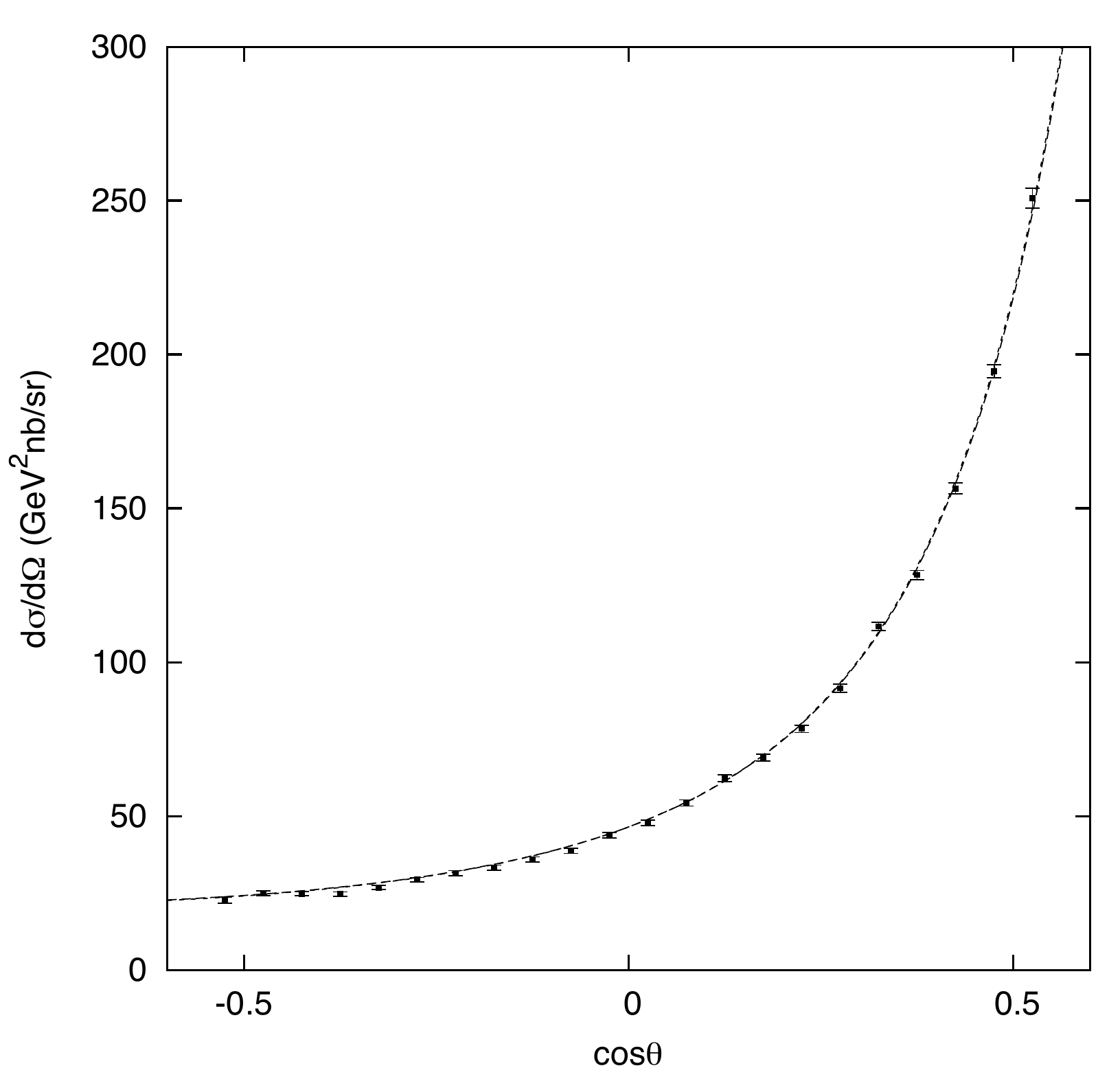} \\
	(a.) \\
	\includegraphics[scale=0.7]{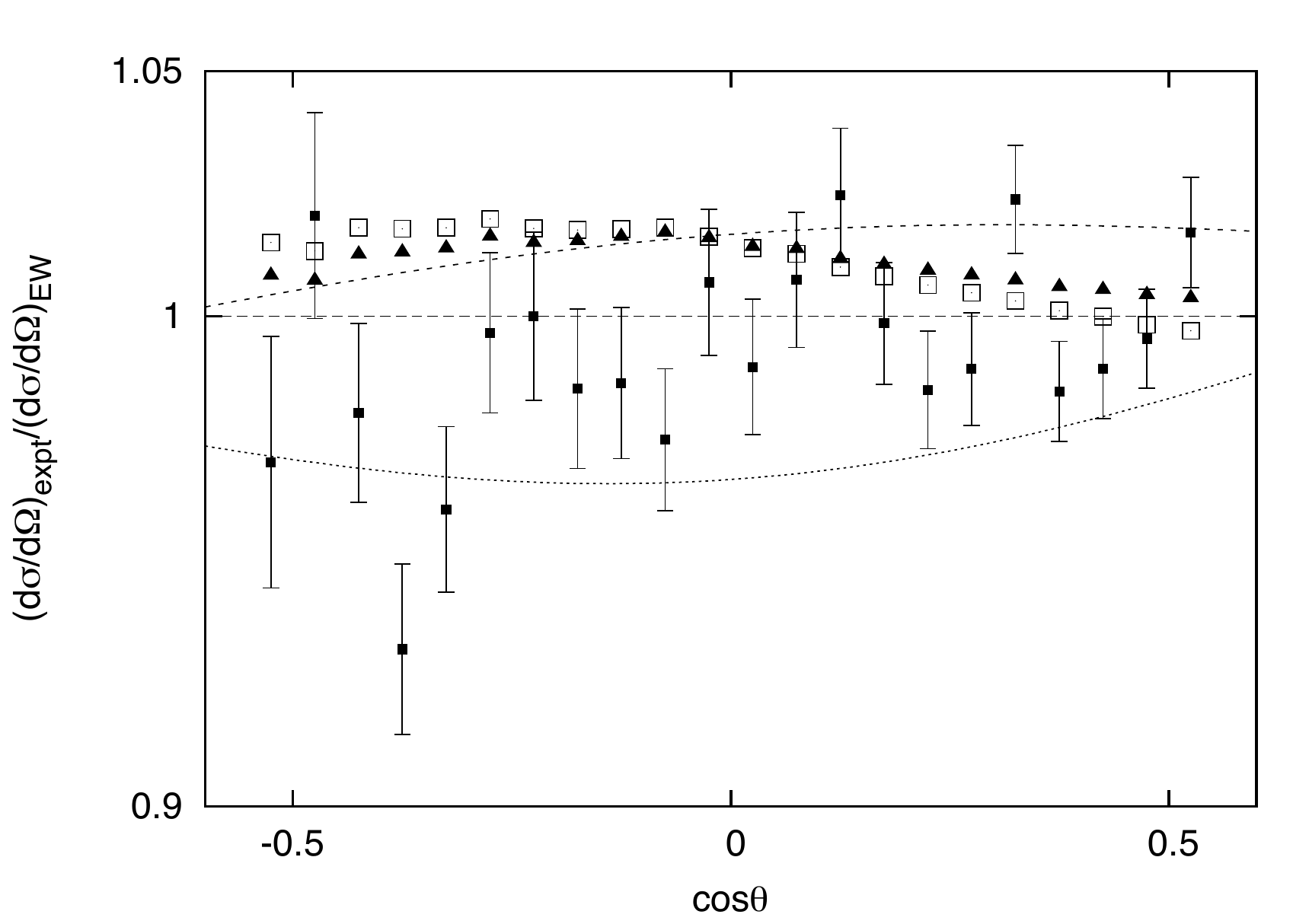} \\
	(b.)
	\end{tabular}
      \end{center}
\caption{\label{bhabha}  The $e^+e^-$ scattering differential cross section $d \sigma / d \Omega$ as a function of $\cos \theta$ at 29 GeV \cite{bhabha}.  (a.) The pure QED cross section, the shape corrected cross section, with $\epsilon = 0.005$, and the data \cite{bhabha} are plotted.  The deviation between the two curves is slight and the curves agree with the data. (b.) The ratio of the experimentally determined differential cross section and the predicted electroweak cross section at lowest order shows the deviations clearly.  The pure QED prediction (empty squares) deviates from the lowest order electroweak result \cite{bhabha}.  The shape-corrected pure QED prediction (solid triangles) more closely matches the electroweak result.  The positive tilt in the corrected Drell parameterization (top and bottom curves) is due to the correction for $\epsilon = 0.005$.}
\end{figure}

The comparison between the model and the data is shown in figure \ref{bhabha}.  
The data from \cite{bhabha}, the pure QED prediction at lowest order, and the short-distance, shape-corrected prediction ($\epsilon = 0.005$) are plotted.  The second plot contains the differential cross section normalized by the electroweak differential cross section at lowest order, as reported in \cite{bhabha}.   The pure QED result is shown and it clearly deviates from the data and the electroweak result.  Adding in the shape-correction effectively adds positive tilt to the results. The top and bottom curves show the approximate 95 \% confidence limits for the QED cutoff parameters $\Lambda_\pm$ \cite{bhabha}, with $\epsilon = 0.005$ shape corrections.  The effect is to tilt the Drell parametrization curves.  As reported in \cite{bhabha} the scale of $\Lambda_\pm$ indicate point-like scattering down to a length scale of approximately $10^{-18}$ m. The shape correction reduces power at small angles and increases power at large angles, as is clear from figure \ref{dist_plot}.

This model may appear to violate energy-momentum conservation. But it does not obviously do so. Even if the granularity of the angle operator were taken into account, say be taking much smaller fluxes, the combinatorics do not suggest a breakdown in energy-momentum conservation since the effects result only in the modification of the distribution of angles. In the context of scattering for instance,  an angle of $\pi$ between the products of a $2 \rightarrow 2$ scattering event is possible. 

This example exhibits the potential accessibility of the shape corrections to observational tests.  Any constraints on $\epsilon$ would require an analysis of the corrections, such as one starting with the QED interaction, and would require disentangling the various normalizations of the data that assume the usual flat spatial geometry.  If the corrections were to affect the results, the shape correction would appear as additional systematic error, such as the tilt shown in figure \ref{bhabha}.

\section{Discussion}
\label{concl}

This paper introduces a new quantum gravity phenomenology, one that explores effects arising from combinatorial structures in the deep spatial quantum geometry of LQG.  The example model used here relies on the combinatorics of a specific discrete model of spatial geometry, that of a single atom of spatial geometry, the spin network vertex. This model shows that potentially observable effects of quantum geometry need not be tied to (obvious) violations of local Lorentz symmetry and that a scale above the fundamental scale of the theory can arise out of combinatorial effects. It was demonstrated that in the context of the atom of space, the underlying combinatorics may be enough to determine corrections to flat, continuum 3-geometry.

The simple model is defined by three assumptions.  First, the spin network node is simple and monochromatic; all the links incident to the node in the lowest representation $j=1/2$.  Second, that the probability measure on the intertwiners is uniform;  in the intertwiner basis of the angle operator, at fixed flux $\vec{s}$ all possible intertwiners are equally likely.  Then, if the angle spectrum is to resemble the angles of continuum, flat 3-geometry, the fluxes must be very large, resulting in an effectively continuous set of intertwiner core labels $\vec{n}$, which determine the angle $\theta$.  Third, a measurement of angle, when it may be traced to a suitably small scale $\ell_s$, is a measurement of the underlying quantum geometry. This is not so radical an assumption for it amounts to the observation that the scalar product of two vectors $\vec{u} \cdot \vec{v} = h_{ij} u^i v^j$ in the context of quantum geometry, depends on the effective local geometry when the process is sufficiently localized.

To the extent that the spatial geometry is described by the assumptions of this model, the model predicts modifications to microscopic angular geometry.  A key step lies in identifying the small parameter depending on the state of the atom of geometry in the spin network model.  This asymmetry or shape parameter $\epsilon = \sqrt{s_1 s_2}/s_3$ is specific to the model and is a measure of the asymmetry of the angular flux or `area' to the background flux or area of the surfaces $S_k$ in dual complex of the node.  In the mixed state given by (\ref{mixedstate}) the probability distribution of polar angles is modified.  The usual $\sin \theta$ distribution is recovered for small values of the parameter $\epsilon$.

By analyzing angular correlation data constraints can be placed on the shape parameter.  Scattering of ``point-like" particles such as in Bhabha scattering could place constraints on the shape parameter $\epsilon$ at the scale $M_s$ set by the center of mass energy.  The analysis in section \ref{scattering} is too simplistic to reach definite conclusions or constraints, but it does show that such effects can be potentially constrained using high energy scattering data. Constraining the deviations form the usual cross sections would specify properties of a ``generic atom of space" via constraints on the parameters.

There are several developments needed before constraints can be placed on the model of an atom of space:
Although the monochromatic assumption is perhaps not too restrictive since these labels are likely generic it seems possible to generalize the counting arguments of \cite{u1} to include the general case of arbitrary spin.  The continuum approximation used could be also checked numerically in the exact, discrete model.

Matter couplings should be introduced.  One way to do this is through local metric corrections to the QED interaction by  smoothing along the lines of \cite{drell}
\begin{equation}
{\cal L'}_\delta (x) = -e \int d^4x \bar{\psi}(x) \gamma_\mu \psi(x)  g^{\mu \nu}_{eff} (x) A_\nu(z)  F_\delta((x-z)^2).
\end{equation}
By expressing the effective metric in terms of combinatoric corrections the expected form of the cross sections could be determined. 

The model is built on assumptions about the atom of 3-geometry. First, the dependence of effective scale on total flux is via the volume operator. However, the spatial volume operator LQG in is not fully understood. There are two volume operator definitions, the Rovelli-Smolin (RS) volume and the Ashtekar-Lewandowski (AL) volume.  The key difference in the analysis of the spectrum is the treatment of embedding information.  Th RS operator does not depend on the embedding of the node. The AL volume has an embedding-dependent sign factor which turns out to strongly effect both the spectrum and the complexity of the analysis. For instance, one interesting result is that the AL volume has no non-vanishing minimum eigenvalue \cite{BR_vol}. Due to the embedding information, the spectrum of the AL volume is not known for high valence vertices.  (See \cite{BR_vol_recent} for recent work on the AL volume spectrum.)  As the role of the embedding information in LQG kinematics is still under debate there are a variety of perspectives on the volume operator.  In this paper I used the scaling property of the RS volume, roughly that the largest eigenvalue scales as $(\sum s_i)^{3/2}$ for volumes large compared to the Planck scale \cite{mikes}. 

Second, determination of angle may occur over a larger subgraph of the network. When an angular measurement is taken, such as in the context of a scattering event, it is not clear that it is possible to distinguish a fundamental spin network from a coarse-grained or effective spin network.  If that is the case then the fundamental graph could be a lower valence graph and the coarse grained sub-graph would be a high valence node.  As the effective length scale of the measurement was increased the graining would become more coarse, total flux would increase and the averaged shape parameter would tend to zero.  Scattering (or other) data give limits on the shape as a function of scale. If the measurement process inherently involved a coarse-graining then the study would be one of a ``molecule" of quantum geometry rather than an ``atom."

Finally, one might suspect that given the large fluxes, the distribution on the space of intertwiner cores $\vec{n}$ would be purely ``statistical" in that it should be given by the distribution of points $\vec{n}$ from the sum over unit vectors with random orientations  and fixed length. This may be seen to be equivalent to a random walk in 3-space. 
The Rayleigh distribution or a ``radial" distance $|\vec{n}|$ covered in $s$ steps of equal length in three spatial dimensions is
\[
P^{(3)}_s(\vec{n}) = \frac{\vec{n}^2}{s^{3/2}} e^{-3 \vec{n}^2/s} \neq p_{\vec{s}} (\vec{n})
\]
The distribution is not equivalent to the one used the model. In the sense that the expressions differ in spatial dimension we can see from this that the resulting combinatorial corrections are not ``statistical".

\begin{ack} 
It is a pleasure to thank the members of the quantum gravity, cosmology, and quantum foundations groups at the Perimeter Institute for insightful discussion. I gratefully acknowledge the generous support of Hamilton College and the Perimeter Institute.
\end{ack}

\appendix
\section{Angle Operator in embedded framework}
\label{embedded}

\begin{figure}
      \begin{center}
      \begin{tabular}{cc}
      \includegraphics[scale=2.1]{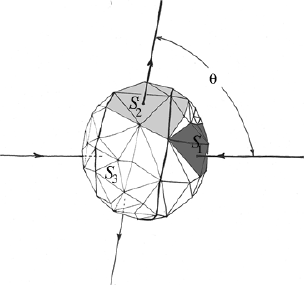} & \includegraphics[scale=0.15]{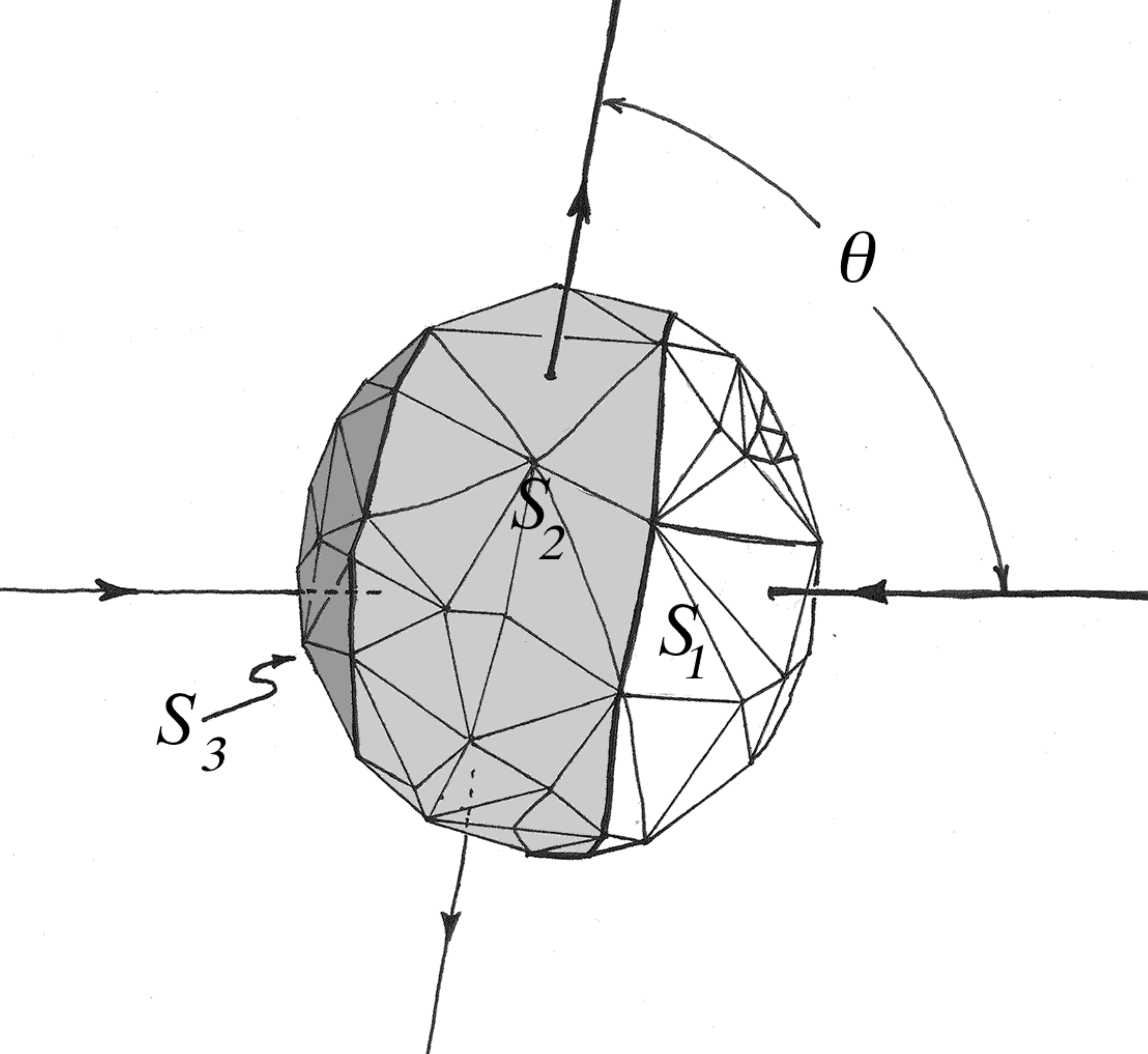}  \\
      (a.) &
	(b.)
	\end{tabular}
      \end{center}
\caption{\label{surfaces3} The three regions $S_k$  in the surface dual to the node.  
(a.) The symmetric identification of surfaces $S_1$ and $S_2$ as done in \cite{angle}. (b.) The polar angle identification of surfaces, with an annular region $S_2$. }
\end{figure}

In the embedded spin network framework the angle operator is defined using a partition of a closed surface around a node. Using the surfaces $S_1, S_2$ shown in figure \ref{surfaces3}, the flux variables $E^i_S = \int_S d^2 \sigma n_a E^{ai}$, and the area operator of the surfaces $A_ {S}$, the angle operator is defined as
\begin{equation}
\theta^{(12)}_{n} := \arccos  
\frac{ E^{i}_{S_{1}} \, E^{i}_{S_{2}} }
{A_ {S_{1}} A_{S_{2}}}.
\end{equation}
All these operators commute. Since the surfaces depend on the graph the operator is explicitly graph dependent. The spectrum is the same as in equation (\ref{angle_spectrum}) if all the links are oriented in the same direction. For arbitrary orientations the operator may be written as
\begin{equation}
	\frac{ L^{2}_{(1_{+}+2_{+})} - L^{2}_{(1_{+}+2_{-})} -
	L^{2}_{(1_{-}+2_{+})} +L^{2}_{(1_{-}+2_{-})} }{ 2 \sqrt{L^{2}_{(1)}}
	\sqrt{L^{2}_{(2)}}}
\end{equation}
in which $1_{+}, 1_{-}$ ($2_{+}, 2_{-}$) label the oriented links oriented outward or inward through the surfaces $S_{1}$ ($S_{2}$), respectively. 

It is clear from the construction that the visualization in terms of the surfaces is heuristic.  While continuum angles are well approximated by fixed semi-classical fluxes, the picture of figure \ref{surfaces3}(b.) suggests that by varying over the regions $S_i$ we could obtain a distribution that is peaked on the appropriate continuum angle $\theta$.  But this is not the case. The continuum angular distribution is obtained at fixed semi-classical fluxes $\vec{s}$.  Of course, in a model without the uniform probability assumption the situation would be different.

The definitions in Ref. \cite{angle} for the scalar product and cosine operators are defined using intersecting surfaces. While these operators are not graph dependent, they do have ordering ambiguities.

\section{Integration of the distribution $P_s(\theta)$}
\label{integration}

For large $s$ the limits on the $n_i$ integrations $(1,s_i)$ may be extended to $(0,\infty)$; the error is $O(1/s_i)$. Re-expressing the delta function in terms of $n_3$ the distribution defined in equation (\ref{rawdisttheta}) becomes
\begin{equation}
P_{\vec{s}}(\theta) = \int_0^\infty d^3n  \, \frac{\delta (n_3 - n_3^* )}{|\partial g(\vec{n},\theta)/\partial n_3|}  \prod_{i=1}^3 \frac{n_i}{s_i} 
\exp \left( -\frac{n_i^2}{2s_i} \right)
\end{equation}
where $g(\vec{n},\theta) := \theta - \arccos \left( ( n_3^2 - n_1^2 - n_2^2) / 2n_1n_2 \right)$ and $n_3^*=\sqrt{n_1^2 + n_2^2 +2 x n_1n_2}$ are its roots, with the usual definition $x=\cos \theta$. Performing the trivial $n_3$ integration gives 
\begin{equation}
P_{\vec{s}}(\theta)  = \sin \theta \int_0^\infty d^2n \frac{(n_1 n_2)^2}{s_1 s_2 s_3} 
\exp \left[- \left( \frac{n_1^2}{2 s'_1} + \frac{n_2^2}{2 s'_2} + \frac{ n_1 n_2 x }{s_3} \right) \right]
\end{equation}
in which  $s'_i := s_i/(1+s_i/s_3) =: s_i \delta_i$, $i=1,2$. For the moment I set $\delta_i=1$, but will comment on these azimuthal asymmetry factors shortly. The next integration is a straightforward quadratic
\begin{equation}
\begin{split}
P_{\vec{s}}(\theta)  = \sin \theta \int_0^\infty dn_1 \left\{ - \frac{\epsilon^2 x n_1^3}{s_1^2} e^{-n_1^2/2s_1}
+ \sqrt{\frac{\pi}{2}} \epsilon \frac{n_1^2}{s_1^{3/2}} \left(1 + (\epsilon x)^2 \frac{n_1^2}{s_1} \right) \right. \\ \left.
\cdot \exp\left[ -\frac{n_1^2}{2s_1} \left(1- (\epsilon x)^2 \right) \right] \Phi\left( \frac{\epsilon x n_1}{\sqrt{2 s_1}} \right) 
\right\}.
\end{split}
\end{equation}
I have introduced the shape parameter $\epsilon := \sqrt{s_1 s_2}/s_3$.  While the first term of the integrand is again a simple quadratic integration, the second is a bit more involved, but still is quadratic.  The result is ($x=\cos \theta$)
\begin{equation}
P_{\vec{s}}(\theta)  =  \sin \theta \, \epsilon \left[  \text{arccot}  ( \epsilon x \sqrt{1 - \epsilon^2 x^2} ) (1+ 2 \epsilon^2 x^2) - 3 \epsilon x \sqrt{1 -\epsilon^2 x^2}   \right] ( 1- \epsilon^2 x^2)^{-5/2}
\end{equation}
At fourth order this expands to 
\begin{equation}
P_{\vec{s}}(\theta)  =  \sin \theta \, \epsilon \left( \frac{\pi}{2} - 4 \epsilon x + \frac{9 \pi}{4} \epsilon^2 x^2 - \frac{29}{3} \epsilon^3 x^3  + \frac{75 \pi}{16} \epsilon^4 x^4 + O(\epsilon^5) \right)
\end{equation}
The actual distribution of angles space, $\rho_\epsilon(\theta) := N P_{\vec{s}}(\theta)$ must be normalized such that the distribution recovers the usual $4 \pi$ solid angle of 3-dimensional spatial geometry in the limit of vanishing $\epsilon$.  Hence, the norm $N$ is fixed by
 \begin{equation}
 \label{normN}
 2 = N \int_0^\pi P_{\vec{s}}(\theta) d\theta,
 \end{equation}
Using the resulting norm and rewriting in terms of Legendre polynomials one finds
\begin{equation}
\label{cgdist}
\rho_\epsilon (\theta) = \sin \theta \left[ 1 - \frac{8}{\pi} P_1(\cos \theta) \epsilon  + \frac{3}{2} P_2(\cos \theta) \epsilon^2 - \frac{2}{5 \pi} \left( P_1(\cos \theta) - \frac{58}{3} P_3(\cos \theta) \right) \epsilon^3 + O(\epsilon^4) \right]
\end{equation}
This is the distribution used in the body of the paper.

Retaining the azimuthal asymmetry factors $\delta_i$ introduced above it is still possible to integrate the distribution $P_s(\theta)$ exactly with the result
\begin{equation} \begin{split}
P_s(\theta) = N \epsilon \sin \theta \left\{ 
(\delta_1  \delta_2)^{3/2} 
 (1 - \delta_1 \delta_2 \epsilon^2 x^2)^{-1/2} (1 - \epsilon^2 x^2 \delta_1 (-3 + \delta_2))
 \arctan \left[ \frac{1 - \delta_1 \delta_2 \epsilon^2 x^2}{\delta_1 \delta_2 \epsilon x}\right] \right. \\ \left.
-3 \epsilon x \delta_1^2  \delta_2^2 \left( 1 - \frac{5}{3} \epsilon^2 x^2 \delta_1 
 (-1 + \delta_2) + \frac{2}{3} \epsilon^4 x^4 \delta_1^2 \delta_2 (-1 + \delta_2) \right)  \right\}
 (1 - \delta_1 \delta_2 \epsilon^2 x^2)^{-2}.
\end{split}
\end{equation}
However upon expanding in $\epsilon$ the  azimuthal asymmetry factors cancel and the result is the same as equation (\ref{cgdist}), as might be expected given the symmetry of the angle operator; there is no parameterization of the azimuthal angle. 

The weights for the averages in equation (\ref{coraveexp}) are
\begin{equation}
\begin{split}
w_1(\theta_o,\delta \theta, \epsilon)  &= \frac{1}{\int_{\Delta \theta} \rho_\epsilon (\theta) d \theta} \int_{\Delta \theta} \rho_\epsilon (\theta) (\theta -\theta_o) d \theta \\
& \simeq \frac{1}{3} \delta \theta^2 \cot (\theta_o)+ \delta \theta^2 \epsilon  \left(\frac{4 \sin (2 \theta_o) \cot
   (\theta_o) \csc (\theta_o)}{3 \pi }-\frac{8 \cos (2
   \theta_o) \csc (\theta_o)}{3 \pi
   }\right)
 \\
w_2(\theta_o,\delta \theta, \epsilon)  &= \frac{1}{\int_{\Delta \theta} \rho_\epsilon (\theta) d \theta} \int_{\Delta \theta} \rho_\epsilon (\theta) (\theta -\theta_o)^2 d \theta \\
& \simeq \frac{\delta \theta^2}{3}
\end{split}
\end{equation}

\end{document}